\documentclass[10pt,conference]{IEEEtran}
\usepackage{authblk}
\usepackage{cite}
\usepackage{amssymb}
\usepackage{mathtools}
\usepackage{textgreek}
\usepackage{eucal}
\usepackage{amsmath}
\usepackage{mathptmx}
\usepackage[acronym,toc,shortcuts]{glossaries}

\usepackage{graphicx}
\usepackage{pgfplots}
\usepackage{tikz}
\usetikzlibrary{patterns}
\usepackage{array}
\usepackage{subfigure}
\usepackage{multicol}
\usepackage{booktabs}

\graphicspath{ {./images/} }
\usepackage[utf8]{inputenc}
\usepackage{comment}
\usepackage[utf8]{inputenc}
\usepackage{multirow}
\usepackage{multicol}
\usepackage{xcolor}
\usepackage[linesnumbered,ruled,vlined]{algorithm2e}
\usepackage{algorithmic}
\newacronym{BS}{BS}{base station}
\newacronym{NOMA}{NOMA}{non-orthogonal multiple access}
\newacronym{PD-NOMA}{PD-NOMA}{power domain non-orthogonal multiple access}
\newacronym{ABC}{ABC}{ambient backscatter communications}
\newacronym{WiFi}{WiFi}{wireless fidelity}
\newacronym{UPWDs}{UPWDs}{user proximity wireless devices}
\newacronym{UPWD}{UPWD}{user proximity wireless device}
\newacronym{UT}{UT}{user terminal}
\newacronym{EMF}{EMF}{electromagnetic field}
\newacronym{AP}{AP}{access point}
\newacronym{SIC}{SIC}{successive interference cancellation}
\newacronym{QoS}{QoS}{quality of service}
\newacronym{RF}{RF}{radio frequency}
\newacronym{OFDMA}{OFDMA}{orthogonal frequency division multiple access}
\newacronym{FCC}{FCC}{Federal Communication Commission}
\newacronym{ML}{ML}{machine learning}
\newacronym{SAR}{SAR}{specific absorption rate}
\newacronym{WHO}{WHO}{World Health Organization}
\newacronym{IARC}{IARC}{International Agency for Research on Cancer}
\newacronym{ICNIRP}{ICNIRP}{International Commission on Non-Ionizing Radiation Protection}

\SetCommentSty{mycommfont}
\SetKwInput{KwInput}{Input}% Set the Input
\SetKwInput{KwInitialize}{Initialize}% set the Output
\SetKwInput{KwOutput}{Output} % set the Output

\begin{document}

\title{\LARGE Green UAV-enabled Internet-of-Things Network with AI-assisted NOMA for Disaster Management}

\author[$\ast$]{Muhammad Ali Jamshed}
\author[$\dag$]{Ferheen Ayaz}
\author[$\dag$]{Aryan Kaushik}
\author[$\star$]{Carlo Fischione}
\author[$\ddag$]{Masood Ur-Rehman}

\affil[$\ast$ ]{College of Science and Engineering, University of Glasgow, Glasgow, U.K}
\affil[$\dag$ ]{School of Engineering \& Informatics, University	of Sussex}
\affil[$\star$]{Division of Network and Systems Engineering, KTH Royal Institute of Technology, Sweden}
\affil[$\ddag$ ]{James Watt School of Engineering, University of Glasgow, Glasgow, U.K}

\affil[ ]{Email: \{muhammadali.jamshed,masood.urrehman\}@glasgow.ac.uk, }
\affil[ ]{\{f.ayaz, aryan.kaushik\}@sussex.ac.uk,}
\affil[ ]{carlofi@kth.se}

%\IEEEoverridecommandlockouts
%\IEEEpubid{\begin{minipage}[t]{\textwidth}\ \\[10pt]
 %       \centering\normalsize{978-1-6654-0086-2/21/\$31.00 \copyright 2021 IEEE}
%\end{minipage}} 

\maketitle
\begin{abstract}
Unmanned aerial vehicle (UAV)-assisted communication is becoming a streamlined technology in providing improved coverage to the internet-of-things (IoT) based devices. Rapid deployment, portability, and flexibility are some of the fundamental characteristics of UAVs, which make them ideal for effectively managing emergency-based IoT applications. This paper studies a UAV-assisted wireless IoT network relying on non-orthogonal multiple access (NOMA) to facilitate uplink connectivity for devices spread over a disaster region. The UAV setup is capable of relaying the information to the cellular base station (BS) using decode and forward relay protocol. By jointly utilizing the concepts of unsupervised machine learning (ML) and solving the resulting non-convex problem, we can maximize the total energy efficiency (EE) of IoT devices spread over a disaster region. Our proposed approach uses a combination of k-medoids and Silhouette analysis to perform resource allocation, whereas, power optimization is performed using iterative methods. In comparison to the exhaustive search method, our proposed scheme solves the EE maximization problem with much lower complexity and at the same time improves the overall energy consumption of the IoT devices. Moreover, in comparison to a modified version of greedy algorithm, our proposed approach improves the total EE of the system by 19\% for a fixed 50k target number of bits.
\end{abstract}

{}

\IEEEpeerreviewmaketitle

\section{Introduction}
\IEEEPARstart{T}{ransportation} and ground infrastructure are often prone to destruction by natural disasters. In 2019, about 27\% of the world's roads and railways were affected by at least one kind of disaster \cite{d1}. Timely warnings and relief operations can play a significant role in mitigating damages caused by disasters, which require effective disaster management and rescue operations. Advanced disaster management systems can potentially utilize emerging 6G technologies and internet-of-things (IoT) networks for communications and control \cite{d2}. Specifically, unmanned aerial vehicles (UAVs) have features such as low weight, small structure, ease of mobility and aerial outreach which are beneficial for surveillance in areas where access to humans or ground nodes is challenging \cite{zhao2018antenna}. For example, UAVs can be used to monitor disaster-affected harsh environments when ground infrastructure is either damaged or prone to destruction \cite{UAV2}. 

The mode of communications and associated technologies are crucially important to provide reliable connectivity during disaster management. It has already been studied that the massive increase in data traffic, high data rate requirements and energy demands are extremely challenging for the practical implementation of 6G IoT networks \cite{9698203}. To cater these requirements, one of the promising 6G technologies is non-orthogonal multiple access (NOMA), known for its higher spectral efficiency (SE), enhanced connectivity, higher cell-edge throughput, and reduction in transmission latency than conventional schemes employing orthogonal multiple access (OMA). Also, NOMA based transmission systems including UAVs are recently emerging as promising solutions for resource allocation and disaster management \cite{CSI1}. Motivated by the applications of NOMA in 6G and UAV-enabled networks, we study a NOMA-based UAV-assisted IoT network in this paper as a disaster management solution.

However, NOMA systems are associated with the trade-off problem between SE and energy efficiency (EE). The increased energy requirement is already a challenge in 6G networks due to the rapid rise in data traffic, network capacity and high frequency band operation \cite{green1}. Furthermore, the additional computational load to support ubiquitous intelligence envisioned in 6G requires high energy consuming machine learning (ML) algorithms. Therefore, green computing and communications are now emerging as potential solutions to reduce energy consumption \cite{green2}. Energy-efficient computing and communication models for green networks are not only helpful to reduce electricity cost but also lead towards a sustainable system suited to future environment goals by balancing demand and supply of energy. EE of relay-based IoT networks and battery-powered devices, such as UAVs, in NOMA systems is being analysed in existing literature \cite{comment4b}. This paper particularly aims to propose a green and sustainable NOMA based UAV-assisted IoT network specifically designed for disaster management. In particular, we study the EE maximization problem of a NOMA-based IoT network including UAVs to provide ubiquitous coverage during a disaster situation. The main contributions of the paper are as follows:

\begin{itemize}
\item We propose a new EE optimization framework to improve the energy utilization of IoT devices available in a disaster region. The proposed framework uses a UAV setup to relay the uplink information of IoT devices using the green artificial intelligence (AI) approach and NOMA-based scheme.

\item We utilize k-medoids, an unsupervised ML technique to cluster the IoT devices, whereas the Silhouette analysis is used to select the best number of IoT devices per subcarrier by exploiting the NOMA characteristics. Furthermore, the power allocation is performed by solving the resulting non-convex problem.

\item We provide a thorough step-wise complexity analysis to validate the enhanced effectiveness of our proposed GREEN-AI algorithm. The computational complexity of the proposed algorithm results lower than the exhaustive search-based optimal method. 

\item We compare our GREEN-AI based proposed solution with a modified greedy algorithm. Simulation results show that the GREEN-AI approach improves the total EE of the system by 19\% for a fixed 50k target number of bits.

\end{itemize}
The rest of the paper is organized as follows. Section II discusses related works, Section III presents the system model and problem definition, Section IV explains the proposed solution, Section V presents simulation findings, and Section VI concludes the paper. 

\section{Related Works}
\subsection{UAVs in Disaster Management}
UAVs can be used for multiple applications in disaster management including monitoring, surveillance, forecasting, early warning predictions and notifications, logistics and evacuation support, search and rescue missions, providing medical aid, and supporting infrastructure and communication systems. In \cite{dis1}, UAVs are used to communicate data related to early warning disaster notifications and act as supporters of resource-constrained wireless sensor networks which play a lead role in sensing geophysical, meteorological or climatic conditions because of their high precision and less operational time. The challenges of UAV-assisted disaster management are discussed in \cite{dis2}, where energy-availability is highlighted as one of the open issues of sustainable UAV systems.
\subsection{Energy Efficiency in UAVs}
Generally, EE is one of the crucial challenges of IoT networks and has been widely discussed in existing literature. EE perspective of UAVs in emergency situations is addressed in \cite{dis3} by simultaneous wireless information and power transfer. A dense deployment of IoT devices is considered and UAVs are equipped with multi-beam antenna arrays for simultaneous transfer of energy to IoT devices. As a future direction, it is suggested to adopt NOMA scheme and model energy optimization as a maximization problem of same-rate. In \cite{CSI1} and \cite{CSI2}, EE is treated as an optimization problem in UAV-enabled NOMA networks but disaster situation is not considered, where IoT devices are often unable to communicate with BS due to infrastructure damage. Apart from NOMA based communication systems, various other solutions related to UAV-enabled applications such as task scheduling and offloading \cite{greedy1}, and trajectory optimization \cite{greedy2} propose to model EE as an optimization problem resolved by simple heuristic algorithms based on greedy approach \cite{greedy3}.
\subsection{Artificial Intelligence (AI) based Energy Efficiency Approaches}
AI based solutions involving Reinforcement Learning (RL) \cite{greedy2} and Deep RL (DRL) \cite{greedy4} combined with greedy algorithms are usually proposed in literature to achieve EE in UAV-assisted applications. In \cite{greedy4}, only the time complexity of a deep neural network during testing phase is analyzed while the computational complexity during learning is not discussed. An energy-efficient and sustainable NOMA-based UAV network is formulated in \cite{DL} by utilizing DRL. However, the complexity analysis of the solution is not evaluated in \cite{DL}. Keeping in view the limitations of related works, the EE solution for NOMA-based UAV networks, particularly for disaster management is worth investigating.

\begin{figure}[!t]
\centering
\includegraphics [scale=0.45]{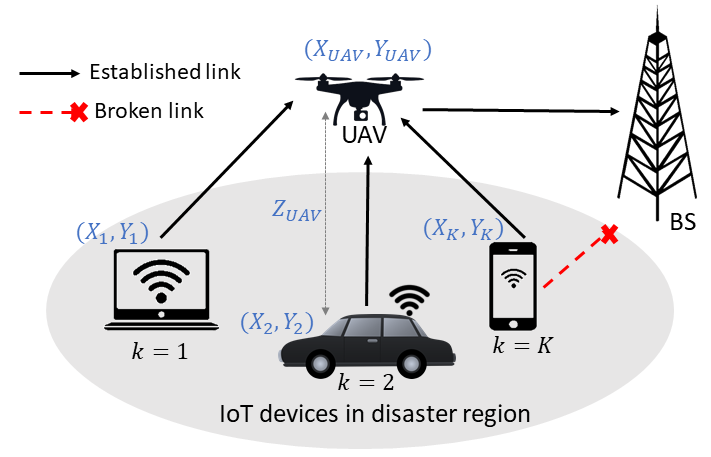}
\caption{UAV-assisted IoT devices in a disaster situation.}
\label{blocky}
\end{figure}
\section{System Model}
%\subsection{Channel Modelling}
Consider an uplink communication scenario for $K$ IoT devices, placed in a disaster region as shown in Fig.~\ref{blocky}. We assume a single cellular-type disaster region, where $K$ deployed IoT devices are unable to communicate directly with the BS. A UAV at a fixed height $Z_{UAV}$ is considered to assist the IoT devices to upload their data to the BS. In the proposed model, we take into account a half-duplex connection in which the NOMA protocol sends data from IoT devices to BS in two time slots. More specifically, the UAV collects the data from $K$ IoT devices using the NOMA protocol, and stores them in the buffer in first time slot. In second time slot, the UAV re-transmits the superimposed signal to BS. The assumptions made for the system are as follows
\begin{itemize}
    \item The system has perfect channel state information (CSI).
    \item A single omnidirectional antenna for uplink communication is assigned with each IoT device.
    \item The channels among devices are independent and suffer from Rician fading.
    \item The position of UAV is static.
    \item We consider the transmission in only first time slot.
\end{itemize}
%\subsubsection{First Time Slot (transmission from IoT devices to UAV)}
The transmission between IoT devices and UAV takes place in the first time slot. Let $g_{k,n}$ denote the channel gain of device $k$ over subcarrier $n$. Incorporating the path loss and fading effect, it can be mathematically defined as
 \begin{align}
g_{k,n}=\frac{\beta_o| h_{k,n}|}{d^{2}_{k}},
\end{align}
\noindent where $\beta_o$ is the channel gain power at a distance of one meter as reference, $h_{k,n}$ is the coefficient of fading and $d_{k}$ is the distance between a UAV and $k^{th}$ device. $d^{2}_{k}$ is defined as
\begin{align}
d^{2}_{k}=\Big[(X_k-X_{UAV})^2+(Y_k-Y_{UAV})^2+Z_{UAV}^2\Big],
\end{align}
\noindent where $X$ and $Y$ denote the x-coordinate and y-coordinate of the locations of UAV and IoT devices respectively. Since we are using NOMA protocol, the UAV receives a superimposed signal and decodes it using traditional SIC. The multiplexing process of multiple devices on a single subcarrier occurs due to NOMA resulting in the following interference on $k^{\text{th}}$ device
\begin{align}
{I}_{k,n}=\sum\limits_{l=1,l\neq k}^{U_{k}} p_{l,n} \cdot g_{l,n},
\label{equ5}
\end{align}
\noindent where $p_{l,n}$ is the transmit power of a device $l$ over subcarrier $n$ and $U_k$ represents the limit of allowable devices on a single subcarrier. The SIC is conducted at the receiving side of signal to perform decoding of multiplexed devices. In NOMA uplink situation, the devices with the best channel gain are demultiplexed first, followed by the decoding of the devices with the poorest channel gain. The multiplexing approach used in NOMA heavily relies on SIC's ability to effectively decode a multiplexed signal. It is possible if the following condition is satisfied:
\begin{align}
(p_{k,n}g_{k,n})/I_{k,n}\geq \zeta,
\label{equ6}
\end{align}
\noindent where $\zeta \geq 1$ is the reference threshold \cite{jamshed2022emission}. The achievable data rate is given as\cite {jamshed2022emission}
\begin{align}
    bt_{k,n} ({\alpha_{k,n}},{p_{k,n}})=w\alpha_{k,n} \log_2\Bigg(1+\frac{p_{k,n}g_{k,n}}{\sigma^2+I_{k,n}}\Bigg),
    \label{equ8a}
    \end{align}
\noindent where $\alpha_{k,n}$ is subcarrier allocation index, $w$ is the subcarrier bandwidth and $\sigma^2$ is the noise variance. Thus the total EE of $K$ devices is defined as
\begin{align}
    EE=B^{\text{Sum}}/P_f+P_t,
    \label{equ8b}
    \end{align}
\noindent where $B^{\text{Sum}}=\sum_{k=1}^{K}\sum_{n=1}^{N}bt_{k,n}$ is the total data rate of $K$ devices over $N$ subcarriers, $Bt_n=\sum_{k=1}^{K}bt_{k,n}$ is the total data rate of $K$ devices over $n^{\text{th}}$ subcarrier, $P_f$ is the circuit power and $P_t$ is the flexible transmit power of $K$ devices \cite{kaushik2020joint, kaushik2020}. %Also, the data rate over a subcarrier $n$ is defined as:
% \begin{align}
%     Bt_{k} = \sum\limits_{k=1}^{K}bt_{k,n}(\boldsymbol{\alpha},\boldsymbol{p})
%     \end{align}
\subsection{Problem Formulation}
We define the problem of maximizing EE as follows
\begin{alignat}{1}
&\  {O}1: \underset{{\boldsymbol{\alpha},\boldsymbol{p}}}{\text{max}} EE\\
\text{s.t.} \\ &\  \mathcal C1: \sum\limits_{k=1}^{K}bt_{k,n}(\boldsymbol{\alpha},\boldsymbol{p})=Bt_n, \forall k \nonumber\\
&\ \mathcal C2: \sum\limits_{k=1} ^{K}\alpha_{k,n}p_{k,n} \leq P_{k}^{\max}  \quad \forall k,\nonumber\\
&\ \mathcal C3: \sum\limits_{k=1}^{k}\alpha_{k,n} \leq U_{k}\quad \forall k, \nonumber\\
&\ \mathcal C4:(p_{k,n}g_{k,n})/I_{k,n}\geq \zeta \quad \forall k, \forall n, \nonumber\\
%&\ \mathcal A6: \sum\limits_{i=1}^2\beta_{i,m}\leq 1, m\in\{1,2\},\nonumber\\
%&\ \mathcal A7: 0\leq\xi_m\leq1, m\in\{1,2\},\label{19}
\end{alignat}
%% need repharasing%
\noindent where the total EE of $K$ devices is defined by $EE(\boldsymbol{\alpha},\boldsymbol{p})$ in objective function ${O}1$. Firstly, the constraint $\mathcal C1$ safeguards the quality of service (QoS) of each device. Secondly, constraint $\mathcal C2$ is associated with the maximum allowable power $P^{\max}_k$ of each device. Thirdly, the constraint $\mathcal C3$ represents the maximum number of devices permitted on a particular subcarrier denoted by $U_k$. Finally, the constraint $C4$ is relevant to the effective implementation of SIC at the receiver end. The binary nature of $\alpha_{k,n}$ and the non-affine nature of constraint $\mathcal C1$ make the problem non-convex \cite{luo2006introduction}. To deal with the binary nature of $\alpha_{k,n}$, a standard relaxation procedure can be used. It can be attained by employing a sequential subcarrier and power allocation strategy which involves subcarrier allocation for a certain fixed power and vice-versa \cite{al2013low}. Furthermore, the CSI is expected to be available via utilizing uplink pilot signals.

%%%%%%%%%%%%%%%%%%%%%%%%%%
% where the objective of $\mathcal P$ is to minimize the total transmit power of AmBC-enhanced NOMA cooperative V2X communication. Constraints $\mathcal A1$ and $\mathcal A2$ ensure the minimum data rate, where $C_{min}$ shows its threshold. Constraints $\mathcal A3$ and $\mathcal A5$ limit the transmit power of BS and RSUs, where $P_{max}$ and $Q_{max}$ are the maximum power that BS and RSU can transmit with at the given time. Constraints $\mathcal A4$ and $\mathcal A6$ describe the power allocation according to NOMA protocol. Constraint $\mathcal A7$ controls the reflection coefficient of backscatter sensors. We can see that the problem $\mathcal P$ is non-convex due to $\mathcal A1$ and $\mathcal A2$, making it very challenging to solve. Thus, we first transform and decouple it into two sub-problems and then employ the sub-gradient method to obtain an efficient solution.

\section{The Proposed Energy Efficient Solution}
This section includes a description of subcarrier allocation, power allocation algorithm, and complexity analysis of the proposed solution. Algorithm 1 describes the detailed procedure of subcarrier and power allocation as the proposed solution.

\subsection{Subcarrier Allocation}
NOMA allows $U_{k}$ devices to be allocated on a single subcarrier with varying channel properties and use the levels of received power for demultiplexing, which results in achieving high SE of a communication system. The capability of the receiver to distinguish among power levels governs subcarrier assignment \cite{jamshed2022emission}. An efficient subcarrier assignment is implemented when devices or users on each subcarrier are grouped effectively. To reach greater effectiveness and a higher probability of convergence than heuristic techniques, an ML technique is used to perform subcarrier assignment by exploiting a clustering algorithm for grouping devices or users \cite{jamshed2021unsupervised}. Two simple and popular clustering algorithms are k-means and k-mediods clustering. The k-means algorithm performs sorting according to the nearest mean value and is therefore highly affected by outliers. The k-mediods algorithm performs sorting around mediod, i.e., a designated centre of the cluster \cite{cluster}. In \cite{cluster}, k-mediods is shown as an algorithm with less number of computations. To achieve lower complexity than the solution proposed in \cite{jamshed2021unsupervised}, we utilize k-medoids based clustering instead of k-means. Furthermore, a Silhouette analysis is proposed instead of the traditional elbow method for predicting the optimum number of clusters. The reason for not using the elbow technique is that it offers a range based on elbow criteria, which creates ambiguity in determining the optimal value of cluster $C$. On the other hand, the Silhouette analysis offers enhanced robustness and eliminates ambiguity.

\subsection{Power Allocation} 
After ML based subcarrier allocation, optimized power allocation of each device needs to be achieved to attain maximum $EE$ of $K$ devices available in a disaster-affected region. Since the constraint $\mathcal C1$ in ${O}1$ is non-convex, (\ref{equ8a}) is transformed to overcome its non-linearity as
%is heuristically transformed into ${O}2$ by drop a decision variable and make an approximation
\begin{align}
p_{k,n}=\frac{(\sigma^2+I_{k,n})(2^{r_{k,n}}-1)}{g_{k,n}},
\label{equ15}
\end{align}
\noindent where, 
\begin{align}
r_{k,n} = \frac{bt_{k,n}}{w\alpha_{k,n}}. 
\end{align}
\noindent The problem ${O1}$ can now be converted into
\begin{align}
&  {O}2: \underset{\boldsymbol{r}}{\text{max}} \hspace{2pt}  \sum\limits_{k=1}^{K} \sum\limits_{n=1}^{N}\frac{\alpha_{k,n} (\sigma^2+I_{k,n})(2^{r_{k,n}}-1)}{g_{k,n}}, \label{eqr}\\ 
& s.t. \nonumber\\&\hspace{5pt} \mathcal C5: \hspace{2pt} w\sum\limits_{n=1}^{N}\alpha_{k,n}r_{k,n}=Bt_{n}, \nonumber \\ 
&\hspace{4pt}\mathcal C6:\hspace{2pt} \sum\limits_{n=1} ^{N}\frac{\alpha_{k,n}(\sigma^2+I_{k,n})(2^{r_{k, n}}-1)}{g_{k,n}} \leq P_k^{\max},\nonumber\\
&\hspace{4pt} \mathcal C7:\frac{\alpha_{k,n} (2^{r_{k,n}}-1)(\sigma^2+I_{k,n})}{I_{k,n}}\geq \zeta. \nonumber\\
%&\hspace{4pt}C_3:\hspace{2pt}\sum\limits_{i=1}^{M}\alpha_{i,e}(t) \leq K \quad \forall t.
\label{equ16}  
\end{align}
\noindent Contrary to ${O}1$, the problem ${O}2$ is convex where the constraints are affine. Therefore, its Lagrangian can be easily obtained and is defined as
\begin{align}
r_{k,n}=\max\Bigg(0, \log_2\chi+\log_2\bigg(\frac{w(g_{k,n})}{\ln(2)(\sigma^2+I_{k,n})}\bigg)\Bigg),
\label{equ24}
\end{align}
\noindent where, $\chi$ is expressed as
\begin{align}
\chi=\frac{\lambda_k^{\star}}{(1-(\mu_{k}^{\star}+\delta_{k,n}^{\star}g_{k,n})/I_{k,n}))},
\label{equ25}
\end{align}
\noindent where $\lambda_k$, $\mu_{k}$ and $\delta_{k,n}$ are associated Lagrange multipliers of $\mathcal C5$, $\mathcal C6$ and $\mathcal C7$. Different iterative solutions can be used to solve (\ref{equ24}), as it is a water-filling based equation \cite{jamshed2021unsupervised}. 

\subsection{Complexity Analysis}
The proposed ML-based NOMA and UAV-assisted method consists of three steps, as defined in Algorithm 1. Step A is responsible for grouping devices while relying on k-medoids and the Silhouette method. The computational complexity in the worst-case scenario of the k-medoids method is $\mathcal{O}(K^2CN)$. On the other hand, the computational complexity of the Silhouette method used within the k-medoids is $\mathcal{O}(K^2)$. Step B is responsible for allocating subcarrier to $K$ users. The computational complexity in the worst-case scenario of Step B is $\mathcal{O}(KN)$. Step C is responsible for finding the optimum rate value to meet the required QoS for $K$ devices. The computational complexity in the worst-case scenario of Step C is $\mathcal{O}((KN)^2)$. Overall the worst-case computational complexity of Algorithm 1, i.e., our proposed solution is $\mathcal{O}((KN)^2)$. If ${O}1$ is directly employed as an exhaustive search technique, the worst-case computational complexity of $\mathcal{O}(2^{KN/2})$ is required which is much higher than the proposed algorithm.
\begin{algorithm}[!h]

\caption{Learning based NOMA and UAV-assisted Energy Efficient IoT Framework for Disaster Management (\textbf{GREEN-AI}).}

\begin{algorithmic}[1]
\STATE \textbf{INPUT} ($K$, $N$, $g_{k,n}$, $\boldsymbol{\alpha}$, $\zeta$, $P_k^{\max}$, $Bt_{n}$, $\sigma^2$, $w$)
%\STATE Set $G_{k,n,t}=g_{k,n,t}/\bar{g}_k$ $\forall k,n,t$, $\bar{g}_k=\sum\limits_{t=1}^{T}\sum\limits_{n=1}^{N}g_{k,n,t}/K/T$;\\
%\FOR{$o = 1,\dots,E$}
\STATE \textbf{Step A:} Allocating devices to a group\\
%\FOR{$t$ = 1 : $T$}
%\FOR{$n$ = 1 : $N$} 
%\STATE \textbf{Step 1:} User grouping and setting $M$\\
\FOR{$C_l$ = 2 : $K-1$} 
\STATE k-medoids based clustering $\mathbf{g}_{n}=[g_{1,n},...,g_{K,n}]$;
\STATE Set $U_{n}=C$ according to Silhouette analysis;\\
\ENDFOR
%\STATE Set ;
%\ENDFOR
%\ENDFOR
\STATE \textbf{Step B:} Allocating subcarrier\\
%\FOR{$t$ = 1 : $T$}
%\FOR{$n$ = 1 : $N$} 
\STATE Use $G_{k,n}=\frac{g_{k,n}}{\hat{g}_k}$ $\forall k,n$,
where $\hat{g}_k$ represents mean of channel gain.
\STATE Use $\text{max}(G_{.,n})$ within each cluster for each device.\\
\STATE Assign $S=\frac{K}{U}$ subcarrier to each device.
%\ENDFOR
%\ENDFOR
\STATE \textbf{Step C:} Optimizing power levels\\
\STATE Set $p_{k,n}$=$P_k^{\max}/S$ to evaluate foremost interference;\\
%\FOR{$o = 1,\dots,E$}
%\FOR{$k$ = 1 : $K$}
%\STATE Estimate \\
%\ENDFOR
\REPEAT
%\WHILE{Convergence}
\FOR{$k$ = 1 : $K$} 
\STATE Estimate $r_{k,n}$ employing water-filling and adhering to the SIC constraint;\\
%\STATE Estimate $p_{u,o,\hat{t}}$ utilizing (8);\\
%\IF{$p_{k,n,t}g_{k,n,t}/I_{k,n,t}< \zeta$}
%\STATE Re-calculate $I_{k,n,t}$ by using (4);\\
%\ELSE
%\STATE Find $r_{k,n,t}$ and $p_{k,n,t}$ that satisfy $p_{k,n,t}g_{k,n,t}/I_{k,n,t}\geq \zeta$ ;\\
%\ENDIF
\STATE Re-evaluate $I_{k,n}$ utilizing (4);\\
\STATE Estimate $EE_k$ utilizing (6);\\
\ENDFOR
\UNTIL \textbf{convergence}
%\ENDWHILE
\STATE Compute $EE$ utilizing ($O2$);\\

\STATE \textbf{OUTPUT} $EE$;
\end{algorithmic}
 \label{AI-1}
\end{algorithm}
\section{Results and Discussions}
 \begin{table}[!t]
 \caption{Simulation parameters\label{tabsim}}
 \centering
 \scalebox{1.2}{
 \begin{tabular}{|c|c|}
 \hline
 \textbf{Parameter} & \textbf{Value} \\
 \hline
 Coverage radius & 500\,m \\
 $Z_{UAV}$ & 100\,m \\
 $K$ & 70 \\
 $w$ & 10\,MHz \\
 $\sigma^2$ & -174\,dBm/Hz \\
 $P_{k}^{max}$ & 0.2\,W \\
 $\zeta$ & 1 \\
 \hline
 \end{tabular}
 }
 \end{table}
In this section, we discuss the MATLAB simulations performed to evaluate the performance of our proposed GREEN-AI as an energy-efficient solution for NOMA based UAV-assisted framework and compare it with the greedy algorithm. The IoT devices are placed uniformly in a disaster region. The simulation parameters used are listed in Table~\ref{tabsim}. To model propagation effects, we assume Rician fading. The path loss models used are defined in \cite{wu2016green}.  

\begin{figure}[t]
\centering
\includegraphics [width=\columnwidth]{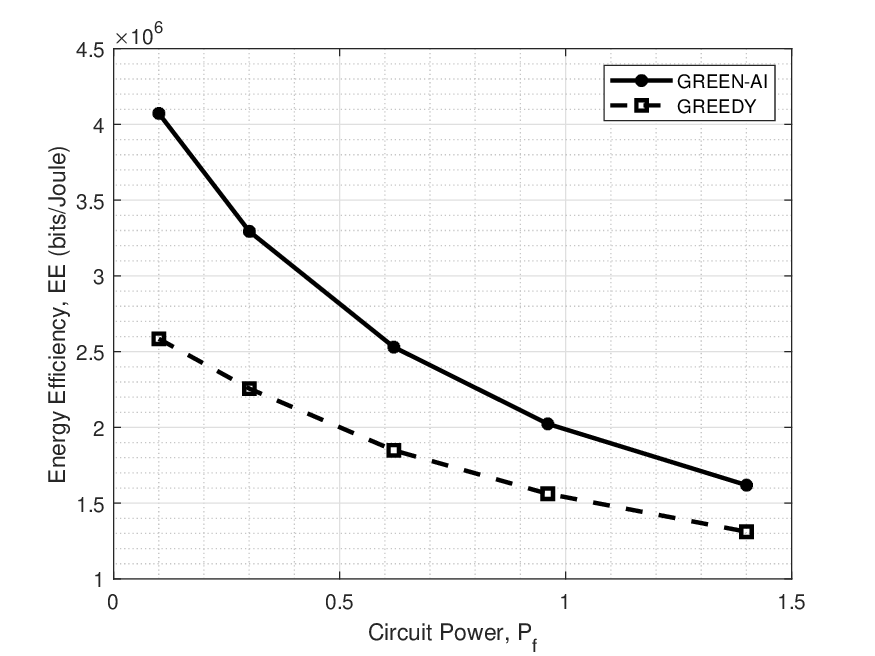}
\caption{$EE$ with respect to $P_f$, when $K=70$, $P_{k}^{max}=0.2$\,W and $Bt_{n}=50$\,kbits.}
\label{fig11}
\vspace{-5.5mm}
\end{figure}

In Fig. \ref{fig11}, we show the effectiveness of the proposed GREEN-AI solution for a fixed uplink power level at each $k^{th}$ device and a fixed target number of bits, $Bt_{n}$. The total $EE$ of the system is observed with respect to $P_f$. In order to perform a fair comparison, we have used the greedy algorithm as a benchmark. The modified version of the greedy algorithm adopted for comparison follows the same steps as defined in Algorithm 1, omitting line 8, which is used to maintain the fairness among all devices. In general, increasing the values of $P_f$ results in a reduction in total $EE$ of the $K$ IoT devices present in the disaster region. In comparison to the modified version of greedy algorithm, our proposed scheme improves $EE$ by 36.5\%, when $P_f=0.1003$, and by 19\%, when $P_f=1.4002$. It is noted that for for a very high circuit power, the performance gap between the two algorithms starts to reduce because higher values of $P_f$ ultimately result in a lower $EE$ of the IoT devices, irrespective of the approach.

%It noted that the high values of $P_{k}^{max}$ result in a low value of $EE$. It is due to fixed target number of bits each value of $P_{k}^{max}$. As shown in Fig. \ref{fig11}, the proposed solution can attain a total $EE$ of 5100 bits/Joule even for a very high circuit power, which shows the stability of our proposed energy efficient solution.

\begin{figure}[t]
\centering
\includegraphics [width=\columnwidth]{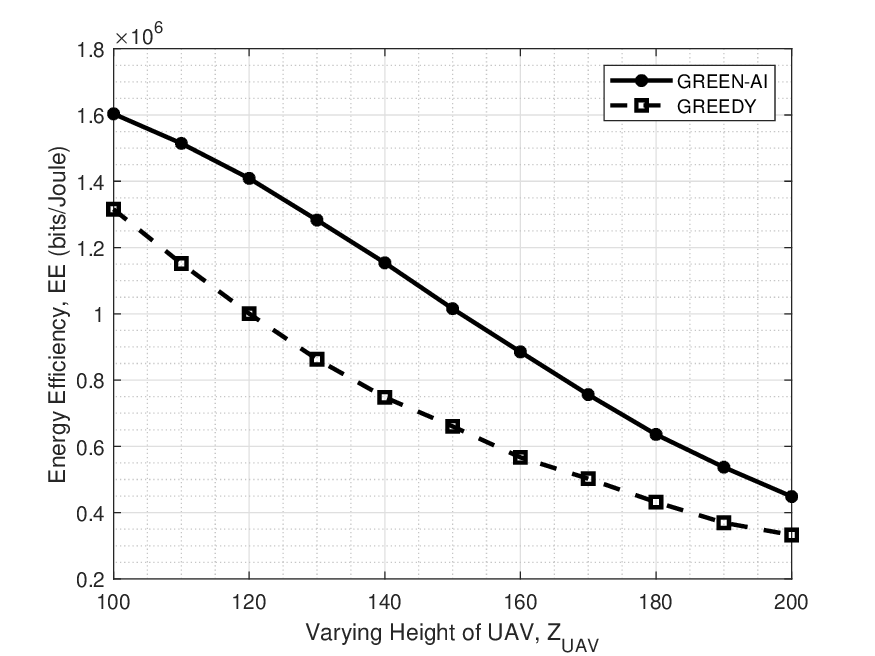}
\caption{$EE$ with respect to $Z_{UAV}$ with $P_f=1.4002$, $P_{k}^{max}=0.2$\,W and $Bt_{n}=50$\,kbits.}
\label{fig4}
\end{figure}

The height of the UAV plays an important role in achieving a suitable amount of total EE of the IoT devices available in the disaster region. In Fig.~\ref{fig4}, we have studied the effectiveness of our proposed algorithm with respect to varying the height of the UAV to adjust the coverage area of the disaster region, whereas target number of bits is fixed at $Bt_{n}=50$\,kbits, and the circuit power is fixed at $P_f=1.4002$. As shown in Fig.~\ref{fig4}, by increasing the height of the UAV the total EE of the IoT devices exponentially decreases. In comparison to the modified version of the greedy algorithm, our GREEN-AI scheme improves the $EE$ by a fair margin and the performance gap remains approximately same with an increase in the height of UAV. The proposed GREEN-AI improves the $EE$ by at least 25\% in comparison of the modified version of greedy algorithm, when the value of $K=70$ and $Z_{UAV}=200$ meters.

\begin{figure}[t]
\centering
\includegraphics [width=\columnwidth]{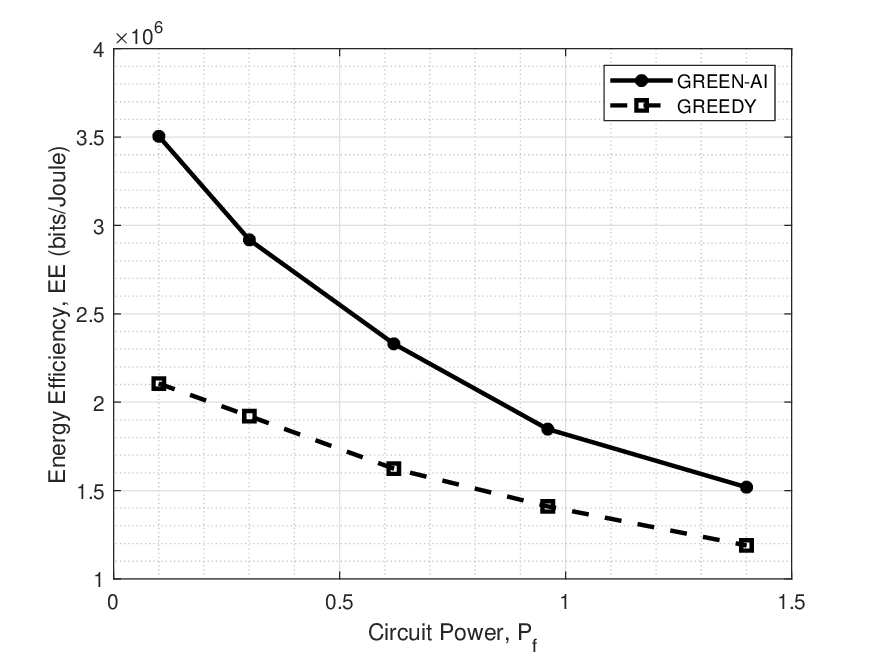}
\caption{{$EE$ with respect to $P_f$, when $K=70$, $P_{k}^{max}=0.2$\,W $Bt_{n}=50$\,kbits, and propagation effects include Rayleigh fading.}}
\label{fig5}
\end{figure}
In order to further validate the stability of our proposed GREEN-AI algorithm, in Fig. \ref{fig5}, we remodel the propagation effects and assume Rayleigh fading for a fixed uplink power level at each $k^{th}$ device and a fixed target number of bits, $Bt_{n}$. In general, an analysis similar to Fig. \ref{fig11} can be drawn, where increasing the values of $P_f$ results in a reduction in total $EE$ of the $K$ IoT devices present in the disaster region. Similarly, for a very high circuit power, the performance gap between the two algorithms starts to reduce because higher values of $P_f$ ultimately result in a lower $EE$ of the IoT devices, irrespective of the approach.

\section{Conclusion}
In this paper, we have presented an energy-efficient NOMA-enabled relay-based UAV-assisted communication model to support the uplink communication of IoT devices situated in a disaster region with the assistance of ML algorithm. Specifically, we have utilized k-medoids to perform resource allocation. The Silhouette analysis is used to find the best number of IoT devices per subcarrier. Finally, the power allocation is performed using iterative methods. Overall, the proposed approach maximizes EE with much lower complexity, in comparison to an exhaustive search. Also, our proposed GREEN-AI scheme improves the total $EE$ of $K$ IoT devices at least by 19\% for a fairly increased value of circuit power as compared to the greedy approach. All the simulation results clearly demonstrate the effectiveness of our proposed UAV-assisted NOMA strategy over the modified baseline greedy algorithm, which is mainly unable to achieve high EE with large number of IoT devices. On one hand, a larger channel capacity and higher spectrum efficiency can be achieved while relying on multiple antenna systems. Also, the increase in the number of antenna elements is linearly proportional to an increase in the number of radio frequency (RF) chains, which greatly impact the EE. Since, our proposed algorithm improves the EE of the entire system, even for an increased amount of circuit power, so therefore, the proposed GREEN-AI algorithm can easily be adopted for multiple antenna scenario. In the future, we aim to further present solutions for multiple UAVs, multiple antennas, and heterogeneous network scenarios. \color{black}
%\vspace{-0.1cm}
% \section*{Acknowledgement}
% This work is supported by the UKRI Higher Education Innovation Fund projects ``Net Zero and Sustainable 6G: Communications, Sensing and Computing," and ``Green and Autonomous UAVs for Advanced Airborne Ecosystem". 

% \ifCLASSOPTIONcaptionsoff
%   \newpage
% \fi

%\bibliographystyle{IEEEtran}
\bibliographystyle{IEEEtran}% This is IEEEtran.bst file
\bibliography{mybib.bib}

\end{document}